\documentclass[floatfix,nofootinbib,superscriptaddress,twocolumn,prl,showpacs,showkeys,preprintnumbers]{revtex4-1}
\usepackage{latexsym,amsmath,amsfonts,amsthm,amssymb,bbm,fdsymbol}
\usepackage{epsfig,graphics,color,calc,graphicx}
\usepackage{subfigure}
\usepackage{mathrsfs}

\newcommand{\bb}[1]{{\mathbb{#1}}}

\newlength{\pecettawidth}
\begin{document}
\title{Blockage induced condensation controlled
by a local reaction}

\author{Emilio N.M.\ Cirillo}
\email{emilio.cirillo@uniroma1.it}
\affiliation{Dipartimento di Scienze di Base e Applicate per l'Ingegneria, 
             Sapienza Universit\`a di Roma, 
             via A.\ Scarpa 16, I--00161, Roma, Italy.}

\author{Matteo Colangeli}
\affiliation{Gran Sasso Science Institute, Viale F. Crispi 7,
00167 L'Aquila, Italy.}
\email{matteo.colangeli@gssi.infn.it}

\author{Adrian Muntean}
\affiliation{Department of Mathematics and Computer Science,
             Karlstad University, Sweden.}
\email{adrian.muntean@kau.se}


\begin{abstract}
We consider the set-up of stationary Zero Range models and discuss the onset of condensation induced by a local blockage on the lattice. We show that the introduction of a local feedback on the hopping rates allows
to control the particle fraction in the condensed phase.
This phenomenon results in a current vs.\ blockage parameter curve 
characterized by two non--analyticity points. 
\end{abstract}

\pacs{}

\keywords{Non-equilibrium stationary states; condensation; blockage effect; Zero Range models.}


\maketitle

The effect of local perturbations of stationary states is 
a fascinating problem in statistical mechanics. At equilibrium, far 
from phase transitions, 
local perturbations typically 
induce local effects, whereas in non-equilibrium stationary states even 
global effects can be observed, for instance on the stationary currents. 

This phenomenon is well known for the totally asymmetric Zero Range process
(ZRP) on the torus 
with time--independent and homogeneous rates
\cite{Evans00,EvansHanney05}, where 
the local perturbation 
(hereafter called \emph{blockage}) 
is the reduction 
of the rate at which a 
single \textit{defect site} of the one--dimensional lattice is updated. 
If the blockage perturbation is small, then no effect persists 
in the large volume limit -- 
computed by keeping constant the ratio between the number of particles and the 
volume of the lattice -- and the macroscopic stationary 
current is unaffected.
Instead, when the perturbation becomes larger,
the current decreases as a result of the condensation of particles 
at the defect site. 
In particular, the current is found to depend non--analytically on 
the intensity of the blockage perturbation.

Condensation phenomena in Zero Range models have been thoroughly 
investigated in the recent literature, cf.\ e.g.\ \cite{GL}, 
where a detailed analysis 
of the literature is provided. 
The effect of the blockage, on the other hand, has also been widely studied 
in the framework of the 
totally asymmetric simple exclusion process
\cite{JanLeb92,JanLeb94,CLST13} and in its parallel 
counterpart \cite{SLM15}. These cases are particularly relevant since the
behavior of the current cannot be explained in terms of the condensation,
due to the imposed exclusion constraint. 
The main issue tackled there was, indeed, to understand whether the decrease of the 
current takes place as soon as the rate on the defect site is 
modified or, alternatively, only when a certain critical value of the 
intensity of the blockage perturbation is reached. 
Related results 
have been also proved in \cite{BSS16}.
It is also worth mentioning that in the recent literature, in 
different framework as non--Markovian process and traffic models, 
ZRP with modified blockage rules have been considered
\cite{HMS2009,HMS2012,CMH,KMH}.

In this paper, we consider the totally asymmetric 
Zero Range model 
and investigate the possibility to compensate the blockage effect 
via a 
local feedback mechanism. This is realized by keeping the rate on the defect 
site constant until 
the occupation number on that site reaches an a priori fixed \emph{activation threshold}. For larger occupation numbers, the rate increases proportionally to the occupation number itself. We show, both numerically and with analytic arguments, that 
such a ``local reaction'' allows to contrast the condensate formation, 
in that it maintains 
the particle fraction in the condensed phase constant for large 
values of the intensity of the blockage perturbation. 
We also point out that, with such a mechanism, 
the current vs.\ blockage intensity curve exhibits two non--analyticity points. 

For the Zero Range models the idea of the activation threshold 
has been introduced in \cite{CCM-MMS16,CCM-JNET16,CCM-CRM16}, where 
different interpretations, ranging from pedestrian dynamics to the 
thermodynamical theory of phase transitions, have been considered. 
As for the pedestrian motion interpretation, the results discussed 
in this paper can be rephrased as follows: particles are regarded as 
pedestrians moving on a lane and the blockage corresponds to the 
presence of a bottleneck or to a lack of visibility (dark, smoke, etc.). 
In this context, particle condensation can be interpreted as 
pedestrian jamming on the blocked spot. The feedback mechanism 
we consider in this paper, see also \cite{CCM-MMS16,CCM-CRM16}, 
can, on the other hand, be interpreted 
as follows: when the number of pedestrian 
on the defect spot exceeds the ``activation threshold"
value, the ability of pedestrians to displace coherently increases thanks to 
information exchange, which becomes significant as soon as the number of 
people on the spot is large enough. 
In this perspective, our results indicate that the jamming 
effect caused by the bottleneck can be compensated by an effective information 
exchange mechanism.

We now define the ZRP to be studied in this Letter, and borrow the notation 
from \cite{EvansHanney05}.
We consider the positive integers $L,N$, 
the finite torus $\Lambda=\{1,\dots,L\}$, and 
the finite \emph{state} or \emph{configuration space}
$\Omega_{L,N}$
made of the states 
$n=(n_1,\dots,n_L)\in\{0,\dots,N\}^\Lambda$
such that 
$\sum_{x=1}^Ln_x=N$.
Given
$n\in\Omega_{L,N}$
the integer $n_x$ is called \emph{number of particle}
at site $x\in\Lambda$ in the \emph{state} or \emph{configuration}
$n$.
The integer $1\le T\le N$ and the real $0<q\le1$ 
are respectively called \emph{activation threshold}
and \emph{blockage parameter}.
Note that for $q$ close to one the intensity of the blockage perturbation 
is small, whereas it is large for $q$ close to zero. 
For any site $x\in\Lambda$,  
the hopping rate $u_x:\bb{N}\to\bb{R}_+$ is defined as follows:
$u_x(0)=0$ for $x=1,\dots,L$, 
$u_1(k)=q$ for $1\le k\le T$ and 
$u_1(k)=q(k-T+1)$ for $T+1\le k\le N$,
and 
$u_x(k)=1$ for $x=2,\dots,L$ and $1\le k\le N$.
The ZRP considered in this context 
is the continuous time Markov process $n(t)\in\Omega_{L,N}$, $t\ge0$,
such that each site 
$x$ is updated with a rate $u_x(n_x(t))$ and, 
once a site $x$ is chosen, a particle is moved 
to the neighboring site $x+1$
(recall that periodic boundary conditions are imposed).

Note also that when $q=1$ and $T=N$ the model reduces to the 
standard Zero Range process whose states can be mapped into 
those of the simple exclusion process.
If $T=N$ and $q<1$ the site at $x=1$ is partially blocked. The effect 
of this kind of blockage is well known, see
\cite[Section~V.1]{Evans00} and \cite[Section~5.2]{EvansHanney05}; here 
we investigate the case $q<1$ and $T<N$ and show that, 
thanks to the local feedback acting on the site $1$, the system is able to 
react to the condensation effect. 

It can be proven, see e. g. \cite[equation~(15)]{EvansHanney05},
that the \emph{invariant} or \emph{stationary measure}
of the ZRP process is 
\begin{equation}
\label{mod020}
\mu_{L,N}(n)
=
\frac{1}{Z_{L,N}}
\times
\Big\{
\begin{array}{ll}
1 & \textrm{if } n_1=0\\
1/[u_1(1)\cdots u_1(n_1)] & \textrm{otherwise}\\
\end{array}
\end{equation}
for any $n\in\Omega_{L,N}$, where the
\emph{partition function}
$Z_{L,N}$ is the normalization constant
\begin{widetext}
\begin{equation}
\label{Zthr}
Z_{L,N}
=\sum_{k=0}^T q^{-k}\binom{L+N-k-2}{N-k}
+\sum_{k=T+1}^{N} 
\frac{q^{-k}}{(k-T+1)!}\binom{L+N-k-2}{N-k} 
\;\;.
\end{equation}
\end{widetext}

The main results discussed in the sequel 
will be deduced in the thermodynamic limit 
$N,L\to\infty$, with $N/L=\rho$ being the 
\emph{global constant density}
and $T/N=\alpha$. 
The use of sistem--size dependent hopping rates, cf. also 
\cite{CG15,GS}, is motivated here by the fact that we want to 
introduce the reaction effect as mildly as possible, in the sense that the 
local rate at site $1$ starts to increase with the number of particles 
only if the local occupation number exceeds an amount proportional to $N$. At the 
end of the paper, we shall also comment on the dramatic effects observed 
if the threshold is chosen independent of $N$. 

The main quantity of interest in our study is the \emph{stationary
current} representing the 
average number of particles crossing a bond between two given
sites in unit time. 
More precisely, since periodic boundary conditions are imposed,
the current does not depend on the chosen bond and is 
given by 
\begin{equation}
\label{mod050}
J_{L,N}
=\mu_{L,N}[u_x]
=
Z_{L,N-1}/Z_{L,N}
\;\;.
\end{equation}
The first equality defines the current, whereas the second one is proven 
in \cite[equation~(11)]{EvansHanney05}. 
Another relevant quantity is  
the stationary 
\emph{particle fraction} at the defect site 
$\nu_{L,N}=\mu_{L,N}[n_1]/N$.
When discussing the thermodynamic limit,  we shall drop the subscripts $L$ and $N$ from the notation 
and write $J$ and $\nu$ for the stationary current and particle 
fraction at site $1$, respectively.


To evaluate the behavior of the partition function in the above 
limit, it is useful to introduce the function 
$I(k)$  by rewriting \eqref{Zthr} as 
$Z_{L,N}=\sum_{k=0}^N\exp\{L I(k)\}$.
To understand where the maxima of $I(k)$ are located, 
we express $I(k+1)-I(k)$ as 
\begin{displaymath}
I(k+1)-I(k)
=
\frac{1}{L}
\Big[
     \log\frac{N-k}{(L+N-k-2)q}
\Big]
\end{displaymath}
for $0\le k\le T-1$ and 
\begin{displaymath}
I(k+1)-I(k)
=
\frac{1}{L}
\Big[
     \log\frac{N-k}{(L+N-k-2)(k-T+2)q}
\Big]
\end{displaymath}
for $T\le k\le N-1$.
By using the two formulas above we can prove that, 
for large $L$, the function $I(k)$ has a single maximum attained 
in $k^*$, with
$k^*=1$ for $q> q_\rho=\rho/(1+\rho)$, 
$k^*=L\lfloor (\rho-q(1+\rho))/(1-q)\rfloor$
for $q_\rho>q>q_\alpha=\rho(1-\alpha)/[1+\rho(1-\alpha)]$,
$k^*=\alpha N$ 
for $q_\alpha>q>q_\alpha/2$, and $k^*$ is given by 
the smallest solution of the equation 
\begin{displaymath}
q[L(1+\rho)-k-2](k-\alpha\rho L+2)=\rho L-k
\end{displaymath}
for $q_\alpha/2>q>0$.
The explicit expression of $k^*$ in the latter case is rather lengthy and will be omitted here. The 
only property we rely on is the fact that, in the thermodynamic limit, 
$k^*/N$ tends to $\alpha$. 

\begin{figure}
\begin{picture}(80,180)(90,0)
\includegraphics[width=0.47\textwidth]{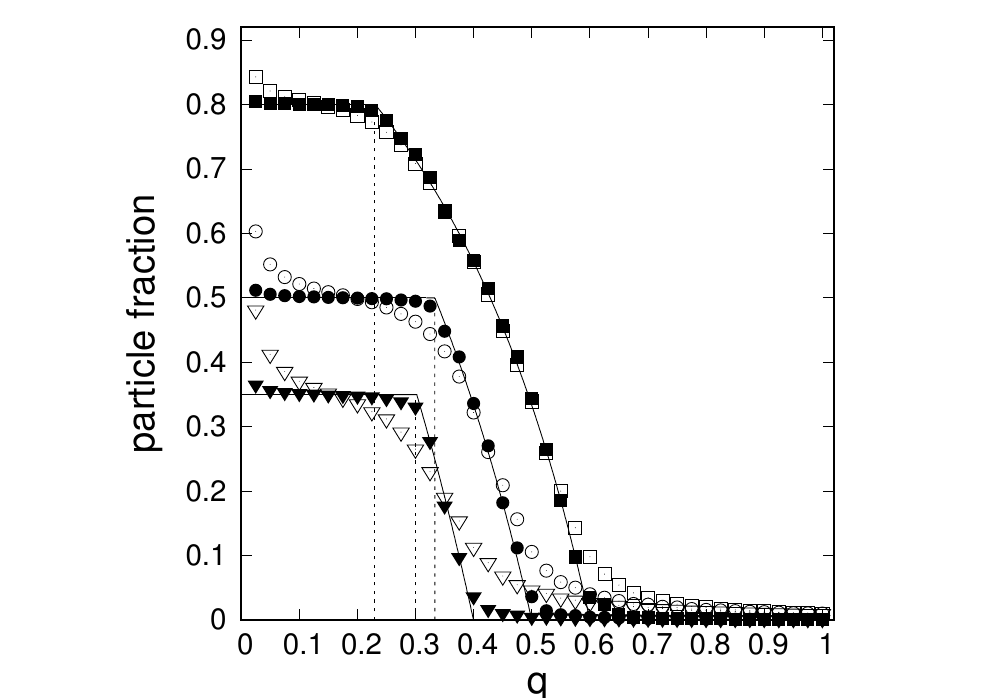}
\end{picture}
\caption{Stationary particle fraction $\nu$ (at the defect site) vs.\ $q$.
Open and solid symbols refer to $L=100$ and $L=1000$ respectively. 
Circles, squares, and triangles refer, respectively, to
$\rho=1$ and $\alpha=0.5$ 
($\smallcircle$ and $\smallblackcircle$, $q_\alpha=0.33$, $q_\rho=0.50$), 
$\rho=1.5$ and $\alpha=0.8$ 
($\smallsquare$ and $\smallblacksquare$, $q_\alpha=0.23$, $q_\rho=0.60$), 
and $\rho=2/3$ and $\alpha=0.35$ ($\smalltriangledown$ and 
$\smallblacktriangledown$, $q_\alpha=0.30$, $q_\rho=0.40$).
Solid lines indicates the analytically predicted behavior in the 
thermodynamic limit. Dotted lines represent the values of $q_\alpha$.}
\label{fig:thr1}
\end{figure}

\begin{figure}
\begin{picture}(80,180)(90,0)
\includegraphics[width=0.47\textwidth]{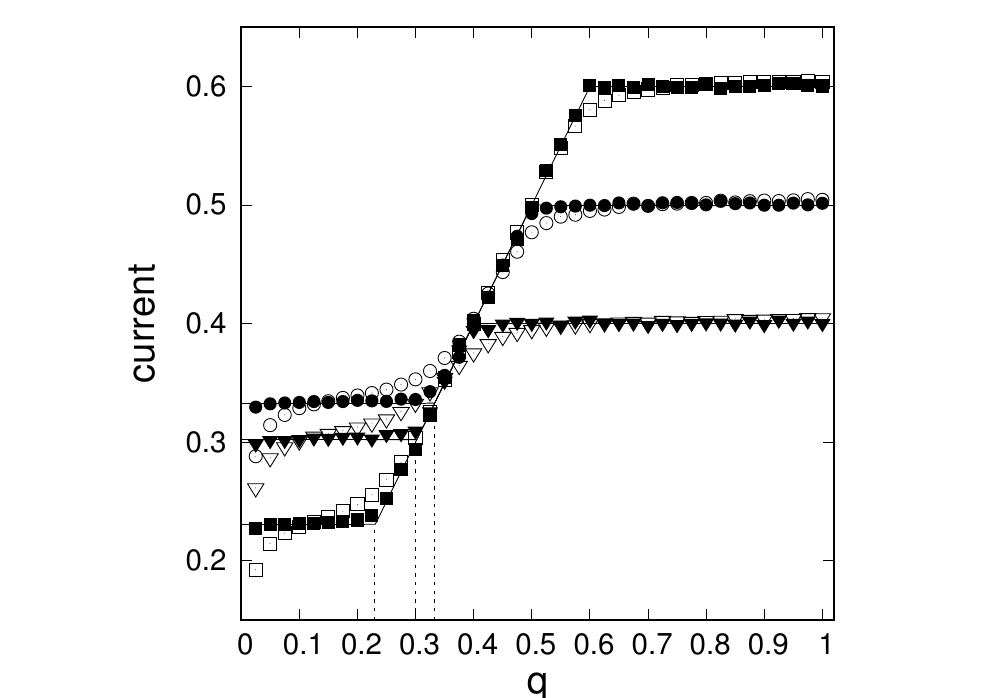}
\end{picture}
\caption{Stationary current vs.\ $q$.
Symbols are as in Figure~\ref{fig:thr1}.
}
\label{fig:thr2}
\end{figure}

The computation of the partition function for the case  $q>q_\alpha$ 
follows the scheme adopted in
\cite[Section~5.2]{EvansHanney05}. 
Indeed, here the terms of the 
second sum in \eqref{Zthr} can be neglected. In particular, for $q>q_\rho$ and $L$ large, 
by expanding the binomials in \eqref{Zthr} for $k\sim O(1)$, 
one finds 
\begin{displaymath}
Z_{L,N}=
\binom{L+N}{N}\frac{1}{1+\rho}\frac{q}{q(1+\rho)-\rho}
\;\;,
\end{displaymath}
which, using \eqref{mod050}, yields 
$J=q_\rho$. Moreover, 
by computing the average occupation number on the defect site, 
one finds 
$\rho/(q(1+\rho)-\rho)$, so that in the thermodynamic limit 
the particle fraction $\nu$ at site $1$ 
vanishes.
Instead, if $q_\alpha<q<q_\rho$, the system undergoes 
condensation. In this case, using the 
Stirling's approximation and computing the resulting Gaussian integral, one finds 
\begin{displaymath}
Z_{L,N}=(1-q)^{-L-1}q^{-N}
\;\;.
\end{displaymath}
Hence, \eqref{mod050} implies $J=q$ and, by computing 
the mean value of $n_1$, one obtains $\nu=1-q/[(1-q)\rho]$.
Thus, in this particular regime,  the particle fraction on the defect site is 
finite (i.e. condensation occurs) and the current is found to decrease linearly when 
$q$ decreases (i.e., the intensity of the blockage perturbation increases).

To treat the case $0<q<q_\alpha$, one has to consider that, for large $L$,
the function $I(k)$ attains its maximum at $\alpha N$. It is then useful to 
rewrite the partition function \eqref{Zthr} by performing the 
changes of variables $h=T-k$ and $h=k-T$ in the first and in the second 
sum, respectively. Then, one can expand the binomials for $h\sim O(1)$ to
find 
\begin{widetext}
\begin{displaymath}
Z_{L,N}
=
\frac{q^{-T}}{[1+\rho(1-\alpha)]^2}
\binom{L+N-T}{N-T}
\Big[
\sum_{h=0}^{T}\lambda^h 
+
\sum_{h=1}^{N-T}
\frac{1}{\lambda^h(h+1)!}
\Big]
\end{displaymath}
\end{widetext}
where we have set $\lambda=q/q_\alpha<1$, so that 
both the two series are converging.
This expression of the partition function allows to compute 
the stationary current via \eqref{mod050}, which leads to 
$J=q_\alpha$. 
Moreover, by computing the mean of the occupation number 
at site $1$ and taking the thermodynamic limit one obtains
$\nu=\alpha$. 

This result is the answer to our initial question: it shows that
the local reaction term affecting the rate at the defect site 
balances, although it does not cancel,
the effect of the blockage 
which originates the condensation. Note that, along this interplay between blockage and local reaction, the phenomenon of condensation is not inhibited: below the critical value 
$q_\alpha$, the particle fraction in the condensed phase stays constant and equal to $\alpha$. 
Moreover, for $0<q<q_\alpha$, the stationary current is also
constant. This means that the behavior of the current versus the 
blockage parameter $q$ reveals two non--analyticity points: one corresponds
to the onset of condensation at $q=q_\rho$, while the second one, at $q=q_\alpha$, points out the 
value of the blockage parameter at which the reaction term becomes 
so effective to stop the rise of the particle fraction in the condensed phase. 

Our analytical results are plotted in Figures~\ref{fig:thr1} and \ref{fig:thr2}
together with the results of Monte Carlo simulations. The model has been 
simulated as follows: call $n(t)$ the
configuration at time $t$, (i)
a number $\tau$ is picked up at random with
exponential distribution of
parameter $\sum_{x=1}^Lu_x(n_x(t))$ and time is update to
$t+\tau$, (ii)
a site is chosen at random on the lattice with probability
$u_x(n_x(t))/\sum_{x=1}^Lu_x(n_x(t))$, and (iii) a particle
is then moved from that site to the neighboring site on the right.
The results shown in the figures reveal a very good match between 
the analytical prediction and the numerical measures. We stress 
that the agreement improves when the lattice size $L$ increases. 
Therefore, the numerical simulations fully confirm 
our description of the main features of the model. 

\begin{figure}
\begin{picture}(80,180)(70,0)
\includegraphics[width=0.47\textwidth]{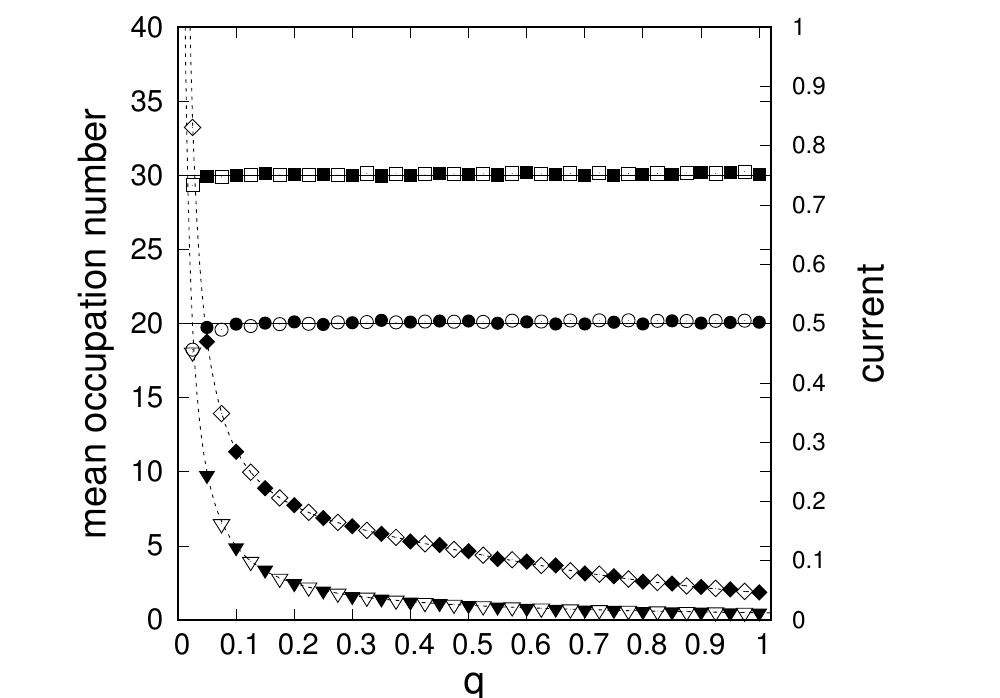}
\end{picture}
\caption{Stationary occupation number at the defect site 
and current vs.\ $q$.
Open and solid symbols refer to $L=100$ and $L=1000$, respectively. 
Circles and squares denote the
current at $\rho=1$ and $T=1$ 
($\smallcircle$ and $\smallblackcircle$, value of the current $0.5$) 
and at $\rho=3$ and $T=5$ 
($\smallsquare$ and $\smallblacksquare$, value of the current $0.75$), 
whereas
triangles and diamonds denote the 
occupation number at site $1$
at $\rho=1$ and $T=1$ ($\smalltriangledown$ and 
$\smallblacktriangledown$)
and at $\rho=3$ and $T=5$ ($\smalldiamond$ and 
$\smallblackdiamond$), respectively.
Solid and dashed lines indicate, respectively, the analytically 
predicted behavior of the current and the occupation number in the 
thermodynamic limit.}
\label{fig:thrT}
\end{figure}

We recall that the threshold in the reaction term has been 
chosen proportional to $N$ to let the reaction effect be weak enough 
(the activation threshold diverges in the thermodynamic limit).
Yet, by setting the threshold equal 
to a constant, the description of the model changes
dramatically. 

The Monte Carlo simulations plotted in Figure~\ref{fig:thrT}
confirm that, in this case, the reaction term 
does inhibit the condensation. Indeed, the plot of 
the current vs.\ the blockage parameter $q$ (scale on the right 
side of the bounding box) is constant, namely, for any value of $q$ 
the current is equal to the one in which no blockage perturbation 
is considered. 
This happens since no condensation is induced in the system, as it 
can be remarked by looking at the plot of the mean occupation number
at the defect site (scale on the left side of the bounding box). 
In such a case, indeed, the mean occupation number at site $1$
is of order one for any value of $q>0$, so that the particle fraction 
$\nu$ tends to zero in the thermodynamic limit. 

This occurs because the local feedback mechanism overwhelms the blockage 
effect and prevents the condensation. 
Indeed, the first sum in \eqref{Zthr} is finite and can be estimated 
by expanding the binomial considering $k\sim O(1)$. 
For the second sum, after performing the change of variables 
$h=k-T$, one observes that, for large $L$, the sum concentrates
on the terms $h\sim O(1)$ and, by accordingly expanding the 
binomials, one finds
\begin{displaymath}
Z_{L,N}
=
\binom{L+N}{N}\frac{1}{(1+\rho)^2}
\Big[
\frac{1-\sigma^{T+1}}{1-\sigma}
+
\sigma^{T-1}(e^{\sigma}-\sigma-1)
\Big]
\end{displaymath}
with $\sigma=q_\rho/q$.
Hence, \eqref{mod050} yields  
$J=q_\rho$. By computing the average occupation 
number at site $1$, for $L$ large and $N=\rho L$, one obtains
\begin{widetext}
\begin{displaymath}
\frac{1}{\binom{L+N}{N}}
Z_{L,N}
(1+\rho)^2
\mu_{L,N}[n_1]
\sim
\frac{\sigma}{(1-\sigma)^2}
[-(T+1)\sigma^T(1-\sigma)+1-\sigma^{T+1}]
+(T-1)\sigma^{T-1}(e^\sigma-\sigma-1)
+\sigma^T(e^\sigma-1)
\end{displaymath}
\end{widetext}
A comparison between numerical data and analytical prediction is 
given in Figure ~\ref{fig:thrT}, where the mean occupation number at site $1$ was used in place
of the particle fraction, because the latter is, in this case, a vanishing quantity. 

\textbf{Acknowledgements}
The authors thank E.\ Presutti, A.\ De Masi, B.\ Scoppola, D.\ Gabrielli and 
C.\ Landim for many discussions and clarifying remarks.


\end{document}